\documentclass[english]{article}
\usepackage[T1]{fontenc}
\usepackage[latin9]{inputenc}
\usepackage{geometry}
\usepackage{color}
\usepackage{array}
\usepackage{float}
\usepackage{textcomp}
\usepackage{amstext}
\usepackage{graphicx}

\makeatletter

\providecommand{\tabularnewline}{\\}
\newcommand{\lyxdot}{.}

\newcommand{\lyxaddress}[1]{
	\par {\raggedright #1
	\vspace{1.4em}
	\noindent\par}
}

\date{}

\makeatother

\usepackage{babel}
\begin{document}
\title{\textbf{Orthogonal-state-based Measurement Device Independent Quantum
Communication}}
\author{Chitra Shukla$^{1,}$\thanks{Emails: chitra.shukla@uni.lu, chitrashukla07@gmail.com},
Abhishek Shukla$^{2}$, Symeon Chatzinotas$^{1}$, Milos Nesladek$^{2}$}
\maketitle

\lyxaddress{$^{1}$Interdisciplinary Centre for Security, Reliability and Trust
(SnT), University of Luxembourg, 1855 Luxembourg, $^{2}$IMO-IMOMEC,
Hasselt University, Wetenschapspark 1, B-3590 Diepenbeek, Belgium.}
\begin{abstract}
We attempt to propose the first orthogonal-state-based protocols of
measurement-device-independent quantum secure direct communication
and quantum dialogue employing single basis, i.e., Bell basis as decoy
qubits for eavesdropping detection. Orthogonal-state-based protocols
are inherently distinct from conventional conjugate-coding protocols,
offering unconditional security derived from the duality and monogamy
of entanglement. Notably, these orthogonal-state-based protocols demonstrate
improved performance over conjugate-coding based protocols under certain
noisy environments, highlighting the significance of selecting the
best basis choice of decoy qubits for secure quantum communication
under collective noise. Furthermore, we rigorously analyze the security
of the proposed protocols against various eavesdropping strategies,
including intercept-and-resend attack, entangle-and-measure attack,
information leakage attack, flip attack, and disturbance or modification
attack. Our findings also show that, with appropriate modifications,
the proposed orthogonal-state-based measurement-device-independent
quantum secure direct communication protocol can be transformed into
orthogonal-state-based measurement-device-independent versions of
quantum key distribution protocols, expanding their applicability.
Our protocols leverage fundamentally distinct resources to close the
security loopholes linked to measurement devices, while also effectively
doubling the distance for secure direct message transmission compared
to traditional quantum communication methods.
\end{abstract}

\section{Introduction\label{sec:Introduction}}

Fully Device-Independence (DI) is a well-known approach which does
not require the learning of devices working principles \cite{Full_DI_1,Full_DI}.
However, it is unfortunate that DI is vulnerable in practice due to
imperfect detectors (require unit efficiency) thereby inviting the
side channel attacks. Moreover, DI requires the Bell (CHSH) inequality
violation to guarantee the security of the protocols. To overcome
these limitations, Lo et al. \cite{Lo_MDI_QKD} introduced the idea
of Measurement-Device-Independence (MDI) which turned out to be a
significant solution to the serious flaws in fully DI, i.e., all the
detector side channel attacks can be avoided along with the advantage
of it doubles the secure communication distance with conventional
lasers as well as it can be implemented with standard optical components
with low detection efficiency (unit efficiency is not required) and
highly lossy channel. Since then the several MDI-QKD have been proposed
in theory and experiment \cite{Experiment_MDI-QSDC}. Further, secure
quantum communication comprises of several important branches, each
addressing specific aspects of cryptography and secure communication
using quantum principles. The most important branches include quantum
key distribution (QKD) \cite{BB84,GV_Protocol}, quantum secret sharing
(QSS) \cite{QSS} and quantum secure direct communication (QSDC) \cite{Long_QSDC1,Long_QSDC2,GV_QSDC,QC},
and quantum dialogue \cite{Ba_An_QD,Our_QD}, QSDC being the most
promising and advance \cite{QSDC_exp1,QSDC_exp2,QSDC_exp3}. QSDC
introduced by Long et al. \cite{Long_QSDC1,Long_QSDC2} is one of
that novel communication technique to send secret messages directly
and securely without the prior generation of quantum keys. Since it
alleviates the requirement of generation and encryption processes,
it is the best solution to share the secret information directly with
low latency. The bottleneck for the implementation of QSDC is unavailability
of quantum memory, which is essential for QSDC protocols. While there
are also memory-free QSDC options available \cite{Memory_free_-QSDC},
recent advancements in quantum memories realizations have optimistically
triggered the research community focus towards QSDC \cite{Memory_QSDC,Memory_MDI_QSDC},
that has further inspired the design and development of DI and MDI
versions of QSDC following the success of DI-QKD \cite{DI_QKD} and
MDI-QKD \cite{Lo_MDI_QKD}. To the best of our knowledge, in this
direction, Long et al. have put forward the first MDI-QSDC protocols
\cite{MDI-QSDC2,MDI-QSDC-LONG} using teleportation and entanglement
swapping. After that several MDI-QSDC protocols have been designed
and proposed in the recent years \cite{Memory_MDI_QSDC,MDI_QSDC_pan,MDI_QSDC_Das,MDI_QSDC_Guo}\textbf{.
}The security of all these protocols relies on the BB84 subroutine
\cite{BB84,Our-OSB}, which implements conjugate coding using two
or more mutually unbiased bases (MUBs), such as ${\left\{ |0\rangle,|1\rangle\right\} }$
and ${\left\{ |+\rangle,|-\rangle\right\} }$, where the security
arrises from Heisenberg-Uncertainty-Principle (HUP). In this approach,
decoy qubits are prepared randomly in non-orthogonal states from either
of these bases to detect potential eavesdropping, while the message
qubits are encoded and decoded using orthogonal states. In contrast
to these traditional protocols, in this work, we aim to explore fundamentally
different resources to design orthogonal-state-based (OSB) MDI-QSDC
and MDI-QD protocols. Specifically, the OSB quantum communication
protocols utilizes single basis (orthogonal states) for encoding-decoding
and eavesdropping checking following GV subroutine \cite{Our-OSB,Kishore-Noise-paper}.
The security framework of OSB protocols stems from the principles
of wave-particle-duality \cite{GV_Protocol} (in the single-particle
scenario) and the monogamy of entanglement \cite{GV_QSDC1} (in the
multi-particle scenario), independent of the need for conjugate coding.
This demonstrates that conjugate coding is not a prerequisite for
achieving secure quantum communication. As a result, OSB protocols
offer significant appeal, particularly from a foundational perspective,
by challenging conventional assumptions about the necessity of conjugate
coding in quantum communication. Interestingly, a significant body
of literature on OSB protocols for secure quantum communication has
emerged in recent years, initiated by some of the authors \cite{Our-OSB}.
This includes protocols for quantum key agreement (QKA) \cite{Our_QKA},
QSDC and deterministic secure quantum communication (DSQC) \cite{Our-OSB,QSDC_Ortho},
as well as quantum dialogue (QD) \cite{Our_QD,AQD}. Further developments
have extended these OSB protocols to semi-quantum settings, including
OSB QKA, QSDC/DSQC, and QD \cite{Semi_QKA_QSDC_QD}, which have even
led to novel applications such as quantum online shopping \cite{Semi_QKA_QSDC_QD}.
However, to date, no attempts have been made to explore OSB protocols
within DI or MDI frameworks. The proposed protocols represent the
first efforts to develop OSB MDI-QSDC and OSB MDI-QD protocols.

Apart from the security aspect, a key motivation for proposing the
OSB MDI-QSDC and OSB MDI-QD protocols is that the decoy Bell qubits
employed form a decoherence-free subspace under collective noise.
Given that collective noise is a major source of decoherence in quantum
communication experiments, identifying decoherence-free states that
can safeguard quantum information from such noise is crucial for reliable
implementation. Interestingly, it has been shown that $|\phi^{\pm}\rangle$
are decoherence free as decoy qubits \cite{Kishore-Noise-paper,Collective_noise}
under collective dephasing noise. The study in \cite{Kishore-Noise-paper}
further demonstrates that when using $|\psi^{\pm}\rangle$ as decoy
qubits, the fidelity reaches unity for phase angles $\varphi=n\pi/2$,
while it drops to zero for $\varphi=(2n+1)\pi/2$. In contrast, the
average fidelity in the BB84 subroutine (uses conjugate coding) does
not exhibit this phase-dependent behavior. Similarly, \cite{Kishore-Noise-paper}
also investigated and revels that $|\psi^{+}\rangle$ and $|\phi^{-}\rangle$
states are decoherence-free subspace under collective rotation noise.
As a result, these states serve as optimal decoy qubits for channels
experiencing collective rotation, making them the best choice for
ensuring security in such environments. In our proposed protocols,
Alice and Bob use $|\psi^{+}\rangle_{d_{1}d_{2}}$ and $|\psi^{+}\rangle_{d_{3}d_{4}}$
states, respectively, as decoy Bell qubits which is decoherence-free
\cite{Kishore-Noise-paper}. Since the security of our protocols relies
heavily on these decoherence-free decoy Bell qubits, it is critical
to protect them from noisy environment. If collective noise is detected
in the communication channel, we can proactively prepare appropriate
decoy qubits (that are decoherence-free) for creating the verification
string necessary for implementing QSB MDI-QSDC and QSB MDI-QD protocols.
In some cases, specific types of noise might even be intentionally
introduced to enhance security, leveraging the known behavior of these
states in noisy environments.

MDI-QSDC protocols are designed to address security vulnerabilities
stemming from imperfections in measurement devices used in quantum
communication, such as side-channel attacks that exploit detector
flaws. By employing Bell state measurements and entanglement swapping,
the protocol shifts the security focus away from the measurement devices,
ensuring that an eavesdropper cannot exploit them. Additionally, MDI-QSDC
enhances communication range by leveraging entanglement swapping,
effectively doubling the distance for secure direct message transmission
compared to traditional methods, without compromising security.

The paper is organized as follows. In Sect. \ref{sec:Introduction},
we have introduced MDI QSDC with contemporary state-of-research and
stated the motivation for OSB MDI-QSDC and MDI-QD protocols. Further
in Sect. \ref{sec:OSB-MDI-QSDC}, we described our OSB MDI-QSDC protocol
step by step, which is followed by our OSB MDI-QD protocol in Sect.
\ref{sec:OSB-MDI-QD}. Subsequently, we analyzed the security of our
protocols in Sect. \ref{sec:Security-Analysis} against most relevant
attacks. 

\section{OSB MDI-QSDC protocol \label{sec:OSB-MDI-QSDC}}

There are three parties in the OSB MDI-QSDC protocol, say Alice, Bob
and Charlie, where Alice (sender) who wants to send her secret messages
to Bob (receiver), and Charlie is an untrusted quantum measurement
device performs Bell measurements and could be fully controlled by
an adversary, Eve. Alice and Bob use one of the following Bell states: 

\begin{equation}
\begin{array}{lcl}
|\psi^{+}\rangle & = & \frac{1}{\sqrt{2}}\left(|00\rangle+|11\rangle\right)\\
|\psi^{-}\rangle & = & \frac{1}{\sqrt{2}}\left(|00\rangle-|11\rangle\right)\\
|\phi^{+}\rangle & = & \frac{1}{\sqrt{2}}\left(|01\rangle+|10\rangle\right)\\
|\phi^{-}\rangle & = & \frac{1}{\sqrt{2}}\left(|01\rangle-|10\rangle\right)
\end{array}\label{eq:Bell_states}
\end{equation}

The following are the steps involved in the protocol.

\textbf{Step1:} \textbf{Preparation:} Alice prepares $n$ number of
a Bell state, i.e., $|\psi^{+}\rangle_{12}^{\otimes n}$. She prepares
the two ordered sequences of all the first qubits as sequence $A_{M}$
(on which she is supposed to encode her secret message later) and
of all the second qubits as sequence $A_{E}$ (which is to be used
for entanglement swapping with Bob's $B_{E}$). Further, she prepares
$n$ number of decoy Bell states $|\psi^{+}\rangle_{d_{1}d_{2}}^{\otimes n}$,
where $d$ stands for the verification qubits as decoy Bell pairs.
She takes $|\psi^{+}\rangle_{d_{1}d_{2}}^{\otimes n/2}$ decoy pairs,
keeps all $d_{1}$ decoys with herself and inserts $d_{2}$ partner
decoys randomly in $A_{E}$ sequence to obtain an extended $A_{E}^{\prime}$
sequence. Similarly, Bob also prepares $n$ number of the Bell states
randomly in $|\psi^{+}\rangle_{34}$ or $|\psi^{-}\rangle_{34}$.
He prepares the two ordered sequences of all the first qubits as sequence
$B_{M}$ and of all the second qubits as sequence $B_{E}$. Subsequently,
he prepares $n$ number of decoy Bell states $|\psi^{+}\rangle_{d_{3}d_{4}}^{\otimes n},$
he takes $|\psi^{+}\rangle_{d_{3}d_{4}}^{\otimes n/2}$ decoy pairs,
keeps all $d_{3}$ decoys with himself and inserts $d_{4}$ partner
decoys randomly in $B_{E}$ sequence to obtain an extended $B_{E}^{\prime}$
sequence. Alice and Bob keep their rest of the decoy Bell pairs $|\psi^{+}\rangle_{d_{1}d_{2}}^{\otimes n/2}$
, $|\psi^{+}\rangle_{d_{3}d_{4}}^{\otimes n/2}$ separately for the
security of the communication of $A_{M}$ and $B_{M}$ sequences to
Charlie. It is to be noted that all decoys $d_{2}$ and $d_{4}$ have
been randomly inserted in the communication channel, i.e., each partner
particle of a decoy Bell pair has random position in extended sequences
$A_{E}^{\prime}$ and $B_{E}^{\prime}$ and the actual sequence is
known to Alice and Bob, respectively.

Here, Alice and Bob can prepare the decoy Bell pairs randomly in any
one (or a random series of all) of the Bell states $\{|\psi^{\pm}\rangle_{dd^{\prime}},|\phi^{\pm}\rangle_{dd^{\prime}}\}$
to keep the decoy state secret, which does not allow Charlie to announce
any particular fake decoy Bell measurement outcome. 
\begin{figure}[h]
\begin{centering}
\includegraphics[scale=0.6]{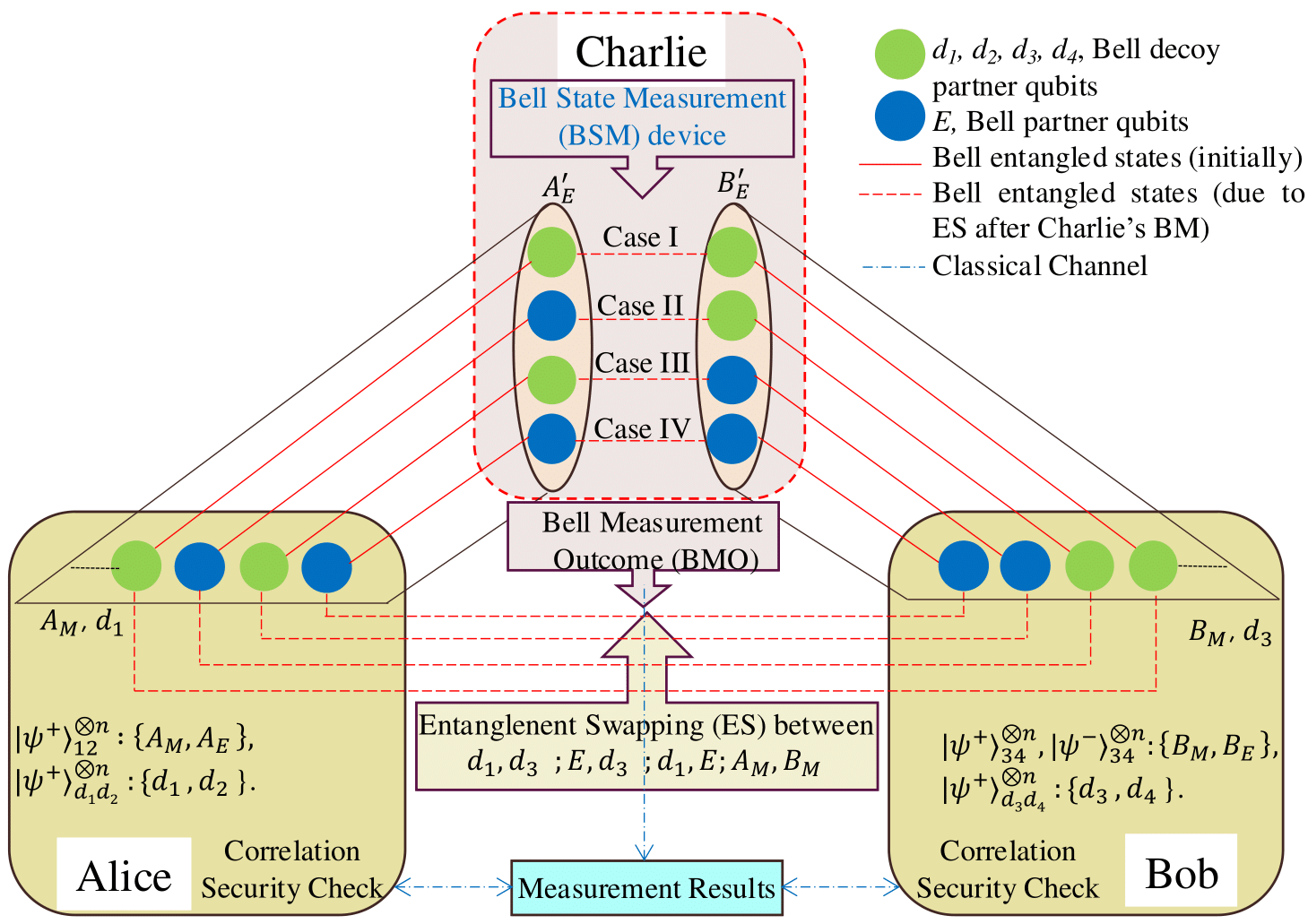}
\par\end{centering}
\caption{(Color online) \label{fig:(Color-online)-Schematic}\textcolor{blue}{A
schematic illustrating the establishment of secure entanglement channel
between Alice and Bob via Charlie's Bell measurement assistance in
our OSB MDI-QSDC protocol.}}
\end{figure}

\textbf{Step2: Transmission of} \textbf{$A_{E}^{\prime}$ and $B_{E}^{\prime}$
sequences}: Alice and Bob keep the sequence $A_{M}$ and $B_{M}$
and their respective decoys $d_{1}$ and $d_{3}$ with themselves
and send the extended $A_{E}^{\prime}$ and $B_{E}^{\prime}$ sequences
to Charlie to allow him to perform the Bell measurement as shown in
Fig \ref{fig:(Color-online)-Schematic}. Without having any knowledge
to distinguish between entangled ($E$) or decoy ($d_{2},d_{4}$)
partner particles in the received sequences $A_{E}^{\prime}$ and
$B_{E}^{\prime}$ from Alice and Bob respectively, Charlie performs
the joint Bell measurement on both the sequences $A_{E}^{\prime}$
and $B_{E}^{\prime}$ and announces the Bell measurement outcome (BMO)
which leads to 4 cases as shown below in Table \ref{tab:4-cases-of}: 

\begin{table}[H]
\begin{centering}
\begin{tabular}{|c|c|c|c|}
\hline 
4 Cases & $A_{E}^{\prime}$ & $B_{E}^{\prime}$  & Actions\tabularnewline
\hline 
Case I & $d_{2}$ & $d_{4}$ & Correlation security check from Eve/Charlie's fake BMO announcement\tabularnewline
\hline 
Case II & $E$ & $d_{4}$ & Correlation security check from Eve/Charlie's fake BMO announcement\tabularnewline
\hline 
Case III & $d_{2}$ & $E$ & Correlation security check from Eve/Charlie's fake BMO announcement\tabularnewline
\hline 
Case IV & $E$ & $E$ & Entanglement swapping (ES) for message transmission\tabularnewline
\hline 
\end{tabular}
\par\end{centering}
\caption{\label{tab:4-cases-of}4 cases of Bell measurement performed by Charlie
and their corresponding actions in the protocol. $E$ indicates the
entangled partner qubit from an entangled pair in $A_{E}^{\prime}$
and $B_{E}^{\prime}$ sequences. $d_{2}$ and $d_{4}$ are partners
of Bell decoys in $A_{E}^{\prime}$ and $B_{E}^{\prime}$, respectively.}

\end{table}

As can be followed in Fig. \ref{fig:(Color-online)-Schematic}, Charlie's
Bell measurement on the sequences $A_{E}^{\prime}$ and $B_{E}^{\prime}$
leads to entanglement swapping, and Alice's and Bob's corresponding
home sequences $A_{M}$, $B_{M}$, along with their decoys $d_{1}$,
$d_{3}$; $E$, $d_{3}$; $d_{1}$, $E$ become pairwise entangled
between Alice and Bob. 

\textbf{Step3: Eavesdropping Check:} After Charlie's BMO announcement,
Alice and Bob announce the positions of decoys $d_{2}$ and $d_{4}$
in their $A_{E}^{\prime}$ and $B_{E}^{\prime}$ sequences, respectively.
They also announce their initially prepared decoy Bell pair state.
For Case I, both Alice and Bob measure their home decoy qubits $d_{1}$
and $d_{3}$ in computational basis $\{|0\rangle,|1\rangle\}$ and
check the correlation according to the Bell measurement outcome of
Charlie's announcement. This is because Alice's and Bob's home decoy
qubits $d_{1}$ and $d_{3}$ are Bell entangled (due to entanglement
swapping) after Charlie performs Bell measurement on decoys $d_{2}$
and $d_{4}$ received from Alice and Bob, let us consider if Charlie's
announces $|\psi^{+}\rangle_{d_{2}d_{4}}$ as his BMO then Alice and
Bob home decoys will be $|\psi^{+}\rangle_{d_{1}d_{3}}$ so if Alice's
measurement (in computational basis) outcome on $d_{1}$ is $|0\rangle(|1\rangle)$
then Bob's outcome on $d_{3}$ will be correlated and should be $|0\rangle(|1\rangle)$.
Further for Case II, Alice measures her corresponding entangled qubit
and Bob measures his home decoy qubits $d_{3}$ in $\{|0\rangle,|1\rangle\}$
and check for correlation according to the BMO of Charlie's announcement.
Similarly, for Case III, Alice measures her home decoy qubits $d_{1}$
and Bob measures his corresponding entangled qubit in $\{|0\rangle,|1\rangle\}$
and check for correlation according to the BMO of Charlie's announcement
(here Bob also has to announce his initial Bell state $|\psi^{+}\rangle_{34}$
or $|\psi^{-}\rangle_{34}$ to help Alice for correlation check).
These three cases ensure the security against an eavesdropper 'Eve'
and any fake announcements (without actually making) Bell measurement
by Charlie, because any eavesdropping/mischievous will disturb the
correlation check that Alice and Bob can easily identify. For example,
if Eve When Alice and Bob calculate the error rate and if they find
the error rate for correlation check below the threshold value then
they continue to the next step, otherwise they discard the protocol
and starts afresh.

It is reasonable to think that case I is enough to ensure the security
against Eve/Charlie, then case II, III can also be used for message
encoding just as case IV and that would make the OSB MDI-QSDC more
efficient. In such a case, Alice just need to announce the initial
Bell decoy state in case III she prepared initially, although in case
II, initial Bell state $|\psi^{+}\rangle_{12}$ is already known. 

\textbf{Step4: Entanglement Swapping:} Now for Case VI, due to Charlie's
Bell measurement on each pair of $A_{E}$ and $B_{E}$ sequences (i.e.,
the qubits $2,4$) leads to quantum entanglement swapping, resulting
the corresponding $A_{M}$ and $B_{M}$ sequences become pair-wise
entangled as Bell states (i.e., the qubits $1,3$) as shown below
in Eq. \ref{eq:Entanglement_Swapping}. 

\begin{equation}
\begin{array}{lcl}
|\psi^{+}\rangle_{12}\otimes|\psi^{+}\rangle_{34} & = & \frac{1}{2}\left(|\psi^{+}\rangle_{13}|\psi^{+}\rangle_{24}+|\phi^{+}\rangle_{13}|\phi^{+}\rangle_{24}+|\phi^{-}\rangle_{13}|\phi^{-}\rangle_{24}+|\psi^{-}\rangle_{13}|\psi^{-}\rangle_{24}\right),\\
|\psi^{+}\rangle_{12}\otimes|\psi^{-}\rangle_{34} & = & \frac{1}{2}\left(|\psi^{+}\rangle_{13}|\psi^{-}\rangle_{24}-|\phi^{+}\rangle_{13}|\phi^{-}\rangle_{24}-|\phi^{-}\rangle_{13}|\phi^{+}\rangle_{24}+|\psi^{-}\rangle_{13}|\psi^{+}\rangle_{24}\right).
\end{array}\label{eq:Entanglement_Swapping}
\end{equation}

It is to be noted, while Bell state discrimination is limited to only
two states in linear optics, non-destructive discrimination of all
four Bell states has been successfully demonstrated with the aid of
ancillary qubits, which can be reused in subsequent measurement iterations
\cite{NDBM1,NDBM2}.

\begin{figure}
\begin{centering}
\includegraphics[scale=0.6]{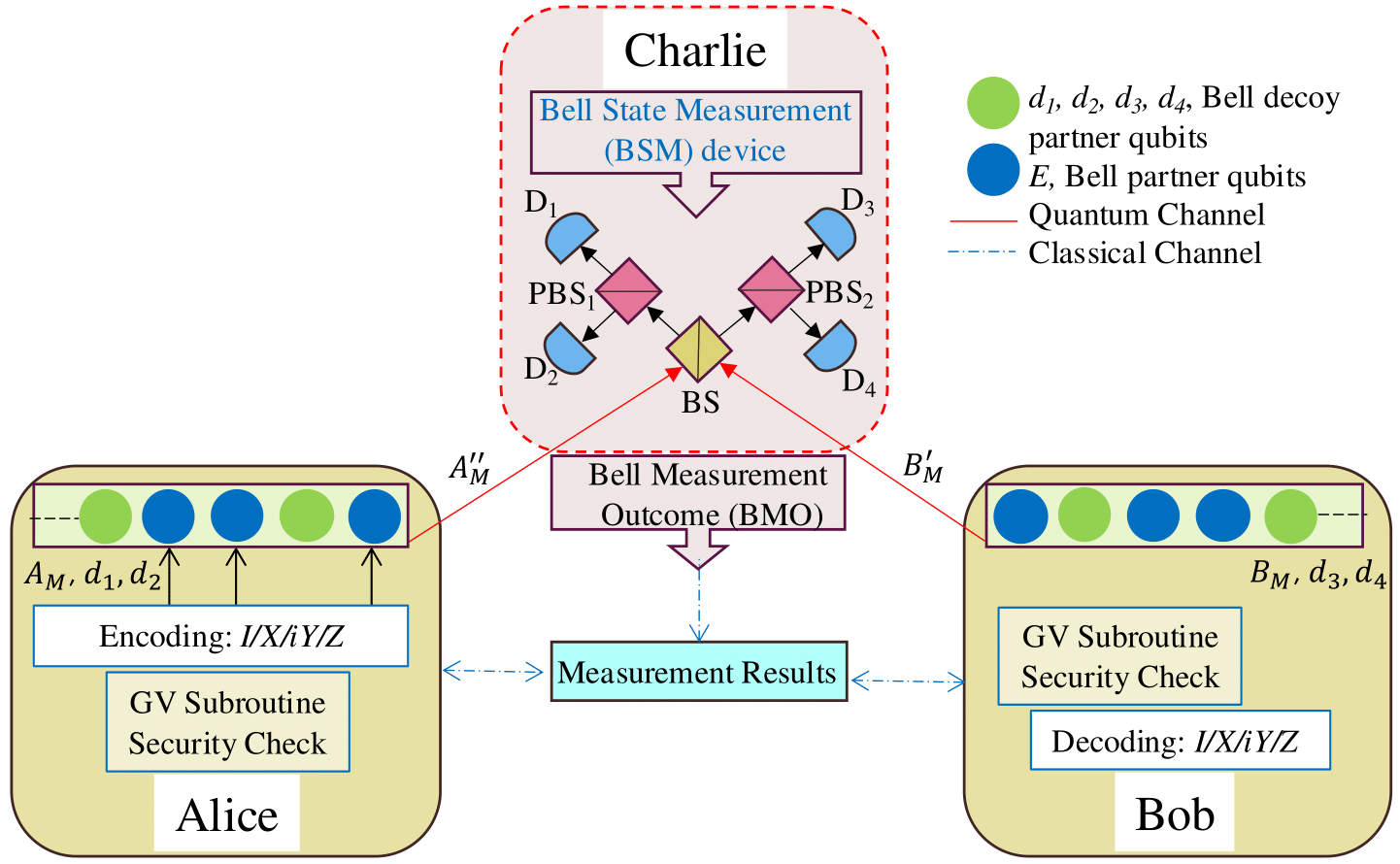}
\par\end{centering}
\caption{(Color Online) \label{fig:(Color-Online)-} \textcolor{blue}{Basic
set-up of Alice and Bob illustrating their cryptographic process (encoding
decoding and eavesdropping checking by GV Subroutine) along with experimental
design of Charlie's Bell measurement device crucial required in our
OSB MDI-QSDC protocol. BS: beam splitter; PBS$_{1}$, PBS$_{2}$:
polarizing beam splitter; $D_{1},$ $D_{2},$ $D_{3},$ and $D_{4}$
detectors.}}

\end{figure}

\textbf{Step5: Message Encoding by Alice: }After the eavesdropping
check is confirmed that there\textbf{ }is no eavesdropping or no fake
announcements from Charlie, Alice (Bob) discarded all the corresponding
partner entangled $(E$) qubits from $A_{M}$$(B_{M})$ which contributed
to cases II (III) and will start message coding process as shown in
Fig. \ref{fig:(Color-Online)-}. Alice holds $A_{M}$ sequence initially
prepared in $|\psi^{+}\rangle_{12}$ state whereas Bob holds $B_{M}$
sequence randomly prepared in $|\psi^{+}\rangle_{34}$ and $|\psi^{-}\rangle_{34}$
states initially. Both the ordered sequences $A_{M}$ and $B_{M}$
are now Bell entangled (i.e., $(A_{M},B_{M})$ pairwise entangled)
due to the entanglement swapping in the last step. Therefore, $(A_{M},B_{M})$
states are only known to Bob based on his initial preparation of $|\psi^{+}\rangle_{34}$
or $|\psi^{-}\rangle_{34}$. Now, Alice applies the unitary (Pauli)
operations $U_{0}=I,$ $U_{1}=X,$ $U_{2}=iY,$ and $U_{3}=Z$ on
the qubits of $A_{M}$ sequence to encode her two-bits of classical
information $00,$ $01,$ $10$ and $11$, respectively. After encoding
her secret messages on $A_{M}$ sequence, Alice finds an encoded sequence
$A_{M}^{\prime}$, in which she inserts rest of the decoy Bell pairs
$|\psi^{+}\rangle_{d_{1}d_{2}}^{\otimes n/2}$ randomly (to follow
the GV subroutine \cite{Our-OSB} for security check in next step)
and sends an extended message encoded sequence $A_{M}^{\prime\prime}$
to Charlie. At the same time, Bob also inserts rest of the decoy Bell
pairs $|\psi^{+}\rangle_{d_{3}d_{4}}^{\otimes n/2}$ randomly (to
follow the GV subroutine \cite{Our-OSB} for security check in next
step) in $B_{M}$ sequence and sends an extended sequence $B_{M}^{\prime}$
to Charlie.

Alice can choose to leave (does not encode) some random qubits of
$A_{M}$ sequence as it is for correlation check in the next step.
This will ensure the security of the message against Charlie. Or Alice
can divide the decoy Bell qubits into two parts to check the security
by GV subroutine using one part and by correlation check using another
part. This strategy would be helpful to trace any bit flip attack
as discussed in subsection \ref{subsec:Flip-Attack:}.

\textbf{Step6: Eavesdropping check:} After receiving the authenticated
acknowledgement of the receipt of all the qubits of $A_{M}^{\prime\prime}$
and $B_{M}^{\prime}$ sequences by Charlie, Alice and Bob announce
the positions of their decoy Bell pairs in their sequences $A_{M}^{\prime\prime}$
and $B_{M}^{\prime}$, respectively. Charlie then measures the decoy
Bell pairs in each sequence $A_{M}^{\prime\prime}$ and $B_{M}^{\prime}$
separately and announces the corresponding BMO. In the absence of
an eavesdropping, Charlie should get each decoy Bell pair in the same
state ($|\psi^{+}\rangle_{dd'}$) as his Bell measurement outcome
as was initially prepared by Alice and Bob, respectively. In between
if Eve tries to eavesdrop and measures the qubits in any of the sequences
$A_{M}^{\prime\prime}$ and $B_{M}^{\prime}$, being unknown of decoys
and message qubits, the decoys Bell pairs will get entanglement swapped
with wrong partner particles and any other Bell measurement outcome
($|\psi^{-}\rangle_{dd'},$$|\phi^{+}\rangle_{dd'}$ or $|\phi^{-}\rangle_{dd'}$)
is the signature of eavesdropping. Alice and Bob compare the decoy
Bell measurement outcomes with that of their initial decoy Bell pair
preparation and calculate the error rate, they also check the correlation
security check on unencoded qubits by Alice, if the error rate is
below the threshold value then they continue to the next step, otherwise
they discard the protocol and starts afresh. 

\textbf{Step7: Decoding of message by Bob:} After being confirmed
there is no eavesdropping, Charlie obtains $A_{M}^{\prime}$ and $B_{M}$
sequences and performs the Bell measurement on $A_{M}^{\prime}$ and
$B_{M}$ sequences and announces his Bell measurement outcomes, using
which Bob can decode the secret message encoded by Alice. This is
so because, only Bob knows which initial Bell state (qubits $1,3$
of column $IV$ of Table \ref{tab:The--sign}) has been shared between
Alice and Bob after entanglement swapping being performed by Charlie
in Step 4, which is in accordance to Bob's initial Bell state preparation
of $|\psi^{+}\rangle_{34}$ or $|\psi^{-}\rangle_{34}$ (column II
of Table \ref{tab:The--sign}).

This is how Alice sends her secret message to Bob via the help of
an untrusted measurement device Charlie, without any information leakage.

\begin{table}
\begin{centering}
\begin{tabular}{|>{\centering}p{0.8cm}|>{\centering}p{2.4cm}|>{\centering}p{2cm}|>{\centering}p{2.6cm}|>{\centering}p{5cm}|>{\centering}p{2.6cm}|}
\hline 
S.No. & Alice \& Bob initial product Bell states $(1,2)$ \& $(3,4)$ & Charlie's Bell measurement result on qubits $2,4$

$(A_{E},B_{E})$ & Alice \& Bob initially shared Bell state after entanglement swapping
(only known to Bob) $(A_{M},B_{M})$ & Charlie's Bell measurement result on message encoded qubits $1,3$
$(A'_{M},B_{M})$ & Bob's decoding of Alice's encoded unitary operation $U_{j}\epsilon\{I,X,iY,Z\},$
where $j\epsilon\{0,1,2,3\}$\tabularnewline
\hline 
1. & $|\psi^{+}\rangle_{12}\otimes|\psi^{+}\rangle_{34}$ & $|\psi^{+}\rangle_{24}$ & $|\psi^{+}\rangle_{13}$ & $|\psi^{+}\rangle_{13}/$$|\phi^{+}\rangle_{13}/$$|\phi^{-}\rangle_{13}/$$|\psi^{-}\rangle_{13}$ & $I/X/iY/Z$\tabularnewline
\cline{3-6} \cline{4-6} \cline{5-6} \cline{6-6} 
 &  & $|\phi^{+}\rangle_{24}$ & $|\phi^{+}\rangle_{13}$ & $|\phi^{+}\rangle_{13}/$$|\psi^{+}\rangle_{13}/$$|\psi^{-}\rangle_{13}/$$|\phi^{-}\rangle_{13}$ & $I/X/iY/Z$\tabularnewline
\cline{3-6} \cline{4-6} \cline{5-6} \cline{6-6} 
 &  & $|\phi^{-}\rangle_{24}$ & $|\phi^{-}\rangle_{13}$ & $|\phi^{-}\rangle_{13}/|\psi^{-}\rangle_{13}/|\psi^{+}\rangle_{13}/|\phi^{+}\rangle_{13}$ & $I/X/iY/Z$\tabularnewline
\cline{3-6} \cline{4-6} \cline{5-6} \cline{6-6} 
 &  & $|\psi^{-}\rangle_{24}$ & $|\psi^{-}\rangle_{13}$ & $|\psi^{-}\rangle_{13}/|\phi^{-}\rangle_{13}/|\phi^{+}\rangle_{13}/|\psi^{+}\rangle_{13}$ & $I/X/iY/Z$\tabularnewline
\hline 
2. & $|\psi^{+}\rangle_{12}\otimes|\psi^{-}\rangle_{34}$ & $|\psi^{-}\rangle_{24}$ & $|\psi^{+}\rangle_{13}$ & $|\psi^{+}\rangle_{13}/$$|\phi^{+}\rangle_{13}/$$|\phi^{-}\rangle_{13}/$$|\psi^{-}\rangle_{13}$ & $I/X/iY/Z$\tabularnewline
\cline{3-6} \cline{4-6} \cline{5-6} \cline{6-6} 
 &  & $|\phi^{-}\rangle_{24}$ & $|\phi^{+}\rangle_{13}$ & $|\phi^{+}\rangle_{13}/$$|\psi^{+}\rangle_{13}/$$|\psi^{-}\rangle_{13}/$$|\phi^{-}\rangle_{13}$ & $I/X/iY/Z$\tabularnewline
\cline{3-6} \cline{4-6} \cline{5-6} \cline{6-6} 
 &  & $|\phi^{+}\rangle_{24}$ & $|\phi^{-}\rangle_{13}$ & $|\phi^{-}\rangle_{13}/|\psi^{-}\rangle_{13}/|\psi^{+}\rangle_{13}/|\phi^{+}\rangle_{13}$ & $I/X/iY/Z$\tabularnewline
\cline{3-6} \cline{4-6} \cline{5-6} \cline{6-6} 
 &  & $|\psi^{+}\rangle_{24}$ & $|\psi^{-}\rangle_{13}$ & $|\psi^{-}\rangle_{13}/|\phi^{-}\rangle_{13}/|\phi^{+}\rangle_{13}/|\psi^{+}\rangle_{13}$ & $I/X/iY/Z$\tabularnewline
\hline 
\end{tabular}
\par\end{centering}
\caption{\label{tab:The--sign}Alice prepares her initial Bell state $|\psi^{+}\rangle_{12}$
and Bob randomly prepares $|\psi^{+}\rangle_{34}$ or $|\psi^{-}\rangle_{34}$.
Accordingly, Column $II$ shows the two possible product states of
Alice and Bob, Column $III$ corresponds to the possible BMO on qubits
(2, 4) by Charlie, and as a result Column $IV$ shows the Bell state
qubits (1, 3) after entanglement swapping (only known to Bob). Column
$V$ and $VI$ show the relationship between Charlie's BMO on Alice's
message encoded qubits and Bob's decoding of unitary operation applied
by Alice.}

\end{table}

Our OSB MDI-QSDC protocol can be adapted into an OSB MDI-QKD scheme
if Alice encodes a random sequence of secret bits, rather than the
meaningful information using operations ($I/X/iY/Z$) to convey specific
messages ($00,$$01,$$10,$$11$). However, converting the reverse\textemdash OSB
MDI-QKD back into OSB MDI-QSDC\textemdash is not possible. Further,
our OSB MDI-QSDC protocol is a unidirectional communication in which
the information flows in one direction of communication (i.e., from
Alice to Bob). It may be interesting to visualize it in bidirectional
communication such that the information flows in both the direction
of communication (i.e., from Alice to Bob and Bob to Alice) simultaneously
using the same quantum channel as happens in the original QD \cite{Ba_An_QD}.

Before we introduce an OSB MDI-QD protocol, it is to be noticed that
the authors introduced an MDI quantum direct dialogue in section 2.3
of \cite{MDI-QSDC2}, where Alice encodes her secret message on her
own travel qubits sequence $M_{A}^{1}$ (for Alice to Bob communication)
but Bob does not encode his secret on the corresponding $M_{B}^{1}$.
Similarly, Bob encodes his secret message on his own travel qubits
sequence $M_{B}^{2}$ (for Bob to Alice communication) but Alice does
not encode her secret on the corresponding $M_{A}^{2}$. Non-availability
of simultaneous encodings of both Alice and Bob on the corresponding
Bell pairs (i.e., on $M_{A}^{1},M_{B}^{1}$ and $M_{A}^{2},M_{B}^{2}$)
while Charlie performs Bell measurement, seems to be different than
the original framework of QD protocols \cite{Ba_An_QD}. Thus, their
MDI quantum direct dialogue rather can be more appropriately described
equivalent to the two QSDCs, first QSDC from Alice to Bob using $M_{A}^{1},M_{B}^{1}$
and second QSDC form Bob to Alice using $M_{A}^{2},M_{B}^{2}$. Because,
when Charlie measures $M_{A}^{1},M_{B}^{1}$ ($M_{A}^{2},M_{B}^{2}$),
it only contains the encoding of Alice (Bob) but not of Bob (Alice).
So when Charlie measures $M_{A}^{1},M_{B}^{1}$ ($M_{A}^{2},M_{B}^{2}$),
both the encodings from Alice and Bob are not available simultaneously
on the same quantum channel, which may not fully align with the requirements
of the original QD framework \cite{Ba_An_QD}.

In principle, in a original QD protocols \cite{Ba_An_QD}, which refers
to the situation, where Alice and Bob send their secret messages to
each other simultaneously on the same quantum channel in which they
both encode their secret messages on the same travel qubit entangled
with another kept as home qubit with Bob. However, we consider a protocol
as QD, when the two different travel qubits which became entangled
(due to entanglement swapping) being encoded by Alice and Bob, respectively.
We emphasize that we can consider such protocols as QD (within original
framework of QD \cite{Ba_An_QD}) even if the two different travel
qubits entangled with each other and each travel qubit carries the
encoding information. The logic behind is that when the two qubits
are entangled then they behave in the same manner regardless of $U_{B}U_{A}$
being encoded only on the second qubit of $|\psi^{+}\rangle_{12}=\frac{|00\rangle_{12}+|11\rangle_{12}}{\sqrt{2}}$
respectively, or $U_{A}$ and $U_{B}$ are separately encoded on first
and second qubit respectively of $|\psi^{+}\rangle_{12}=\frac{|00\rangle_{12}+|11\rangle_{12}}{\sqrt{2}}$.
Following the latter trick, we introduce an OSB MDI-QD protocol in
the next section \ref{sec:OSB-MDI-QD}. The only criteria to be within
QD is that the combined Bell state should be measured by Charlie only
after Alice and Bob apply the operations $U_{A}$ and $U_{B}$ respectively
on their travel qubits. Hence, both the encodings $U_{A}$ and $U_{B}$
are available simultaneously on the same quantum channel. It is to
be noted that in a two-party QD protocol \cite{Ba_An_QD}, it is preferred
to apply $U_{B}U_{A}$ only on the second travel qubit because there
are only two parties and one party Bob starts the protocol and receives
the travel qubit back himself encoded by Alice. But here in OSB MDI-QD
there are three parties where two authorized parties Alice and Bob
separately sending their encoded travel qubits (which become entangled
after entanglement swapping in Step 4) in Step 5 to an unauthorized
party Charlie for Bell measurement. Here, the encoded sequences have
been measured by Charlie only after Alice and Bob send their sequences
$A_{M}$ and $B_{M}$ to Charlie after encoding their secret messages.
With this motivation, now we introduce our OSB MDI-QD protocol in
the next section. 

\section{OSB MDI-QD protocol \label{sec:OSB-MDI-QD}}

QD protocol is a two-way communication scheme that allows both parties,
Alice and Bob, to simultaneously exchange the secret messages on the
same quantum channel utilizing a Bell pair \cite{Ba_An_QD,Our_QD}.
The key advantage of QD is its ability to achieve secure bidirectional
communication in a single session. The proposed OSB MDI-QD protocol
steps could be outlined as follows:

\textbf{Step1:} \textbf{Preparation:} Alice prepares $n$ number of
the Bell states, i.e., $|\psi^{+}\rangle_{12}^{\otimes n}$. She prepares
the two ordered sequences of all the first qubits as sequence $A_{M}$
(on which she is supposed to encode her secret message later) and
of all the second qubits as sequence $A_{E}$ (which is to be used
for entanglement swapping with Bob\textquoteright s $B_{E}$). Further,
she prepares $n$ number of decoy Bell states $|\psi^{+}\rangle_{d_{1}d_{2}}^{\otimes n}$,
where $d$ stands for the verification qubits as decoy Bell pairs.
She takes $|\psi^{+}\rangle_{d_{1}d_{2}}^{\otimes n/2}$ decoy pairs,
keeps all $d_{1}$ decoys with herself and inserts $d_{2}$ partner
decoys randomly in $A_{E}$ sequence to obtain an extended $A_{E}^{\prime}$
sequence. Similarly, Bob also prepares $n$ number of the Bell states
randomly in $|\psi^{+}\rangle_{34}$ or $|\psi^{-}\rangle_{34}$.
Further, he performs the OSB MDI-QSDC protocol as shown in Sect. \ref{sec:OSB-MDI-QSDC}
to securely share the exact information of the prepared Bell state.
However, Alice doesn't have to do so because her initial Bell state
is fixed as $|\psi^{+}\rangle_{12}$ and is a public knowledge. Now,
he prepares the two ordered sequences of all the first qubits as sequence
$B_{M}$ and of all the second qubits as sequence $B_{E}$. Subsequently,
he prepares $n$ number of decoy Bell states $|\psi^{+}\rangle_{d_{3}d_{4}}^{\otimes n},$
he takes $|\psi^{+}\rangle_{d_{3}d_{4}}^{\otimes n/2}$ decoy pairs,
keeps all $d_{3}$ decoys with himself and inserts $d_{4}$ partner
decoys randomly in $B_{E}$ sequence to obtain an extended $B_{E}^{\prime}$
sequence. Alice and Bob keep their rest of the decoy Bell pairs $|\psi^{+}\rangle_{d_{1}d_{2}}^{\otimes n/2}$
, $|\psi^{+}\rangle_{d_{3}d_{4}}^{\otimes n/2}$ separately for the
security of the communication of $A_{M}$ and $B_{M}$ sequences to
Charlie. It is to be noted that all decoys $d_{2}$ and $d_{4}$ have
been randomly inserted in the communication channel, i.e., each partner
particle of a decoy Bell pair has random position in extended sequences
$A_{E}^{\prime}$ and $B_{E}^{\prime}$ and the actual sequence is
known to Alice and Bob, respectively.

Similar to Step1 of OSB MDI-QSDC, Alice and Bob can randomly prepare
the decoy Bell pairs in any of the Bell states $\{|\psi^{\pm}\rangle_{dd^{\prime}},|\phi^{\pm}\rangle_{dd^{\prime}}\}$,
keeping the decoy state secret and preventing Charlie from announcing
a fake Bell measurement outcome.

\textbf{Step2:} \textbf{Transmission of} \textbf{$A_{E}^{\prime}$
and $B_{E}^{\prime}$ sequences}: Alice and Bob keep the sequence
$A_{M}$ and $B_{M}$ with themselves and send the extended $A_{E}^{\prime}$
and $B_{E}^{\prime}$ sequences to Charlie to allow him to perform
the Bell measurement that leads to 4 cases as shown in Table \ref{tab:4-cases-of}.

\textbf{Step3: Eavesdropping Check:} Same as Step 3 of OSB MDI-QSDC
described in Sect. \ref{sec:OSB-MDI-QSDC}. 

\textbf{Step4: Entanglement Swapping: }Same as Step 4 of OSB MDI-QSDC
described in Sect. \ref{sec:OSB-MDI-QSDC}. 

\textbf{Step5: Message Encoding by Alice and Bob: }Once it is confirmed
that there\textbf{ }is no eavesdropping or no fake announcements from
Charlie, Alice (Bob) discards all the corresponding partner entangled
$(E$) qubits from $A_{M}$$(B_{M})$ which contributed to cases II
(III) (they don't discard if they decide to use these cases for message
coding). After that Alice and Bob start message coding process where
Alice holds $A_{M}$ sequence initially prepared in $|\psi^{+}\rangle_{12}$
state whereas Bob holds $B_{M}$ sequence randomly prepared in $|\psi^{+}\rangle_{34}$
or $|\psi^{-}\rangle_{34}$ state initially. Both the ordered sequences
$A_{M}$ and $B_{M}$ are now Bell entangled (i.e., $(A_{M},B_{M})$
pairwise entangled) due to the entanglement swapping in the last step.
Hence, after the Bell measurement announcement of $(A_{E},B_{E})$
from Charlie as shown in column $III$ of Table \ref{tab:The--sign},
Alice and Bob both know the initial Bell state $A_{M},B_{M}$ they
share as shown in $IV$ column of Table \ref{tab:The--sign} according
to Eq. \ref{eq:Entanglement_Swapping}. Unlike OSB MDI-QSDC, here
Alice also know initial Bell state $A_{M},B_{M}$ because Bob has
already executed OSB MDI-QSDC in Step 1 to securely share the exact
information of the prepared Bell state ($|\psi^{+}\rangle_{34}$ or
$|\psi^{-}\rangle_{34}$). Now, Alice applies one of the unitary operations
$U_{0}=I,$ $U_{1}=X,$ $U_{2}=iY,$ and $U_{3}=Z$ on the qubits
of $A_{M}$ sequence to encode her 2-bits of classical information
$00,$ $01,$ $10$ and $11$, respectively. After encoding her secret
messages on $A_{M}$ sequence, Alice finds an encoded sequence $A_{M}^{\prime}$,
in which she inserts rest of the decoy Bell pairs $|\psi^{+}\rangle_{d_{1}d_{2}}^{\otimes n/2}$
randomly and sends an extended message encoded sequence $A_{M}^{\prime\prime}$
to Charlie. At the same time, Bob also randomly applies one of the
unitary operations $U_{0}=I,$ $U_{1}=X,$ $U_{2}=iY,$ and $U_{3}=Z$
on the qubits of $B_{M}$ sequence according to his secret message
to encode his 2-bits of classical information and gets his encoded
sequence $B_{M}^{\prime}$, then he also inserts rest of the decoy
Bell pairs $|\psi^{+}\rangle_{d_{3}d_{4}}^{\otimes n/2}$ randomly
in $B_{M}^{\prime\prime}$ sequence and sends an extended sequence
$B_{M}^{\prime\prime}$ to Charlie.

\textbf{Step6: Eavesdropping check:} This step is same as Step 6 of
OSB MDI-QSDC described in Sect. \ref{sec:OSB-MDI-QSDC}, with the
only difference being that in this case, the sequences $A_{M}^{\prime\prime}$
and $B_{M}^{\prime\prime}$, obtained from the previous step, will
now be checked.

\textbf{Step7: Decoding of message by Alice and Bob:} After being
confirmed there is no eavesdropping, Charlie performs the Bell measurement
on $A_{M}^{\prime}$,$B_{M}^{\prime}$ message encoded sequences and
announces his final Bell measurement outcomes, using which Alice and
Bob can decode the secret messages encoded by Bob and Alice, respectively.
This is so because, Alice (Bob) knows Bob's (Alice's) initial Bell
state prepared in Step 1 for which Bob executed OSB MDI-QSDC as shown
in Sect. \ref{sec:OSB-MDI-QSDC} to securely share the information
about which Bell state ($|\psi^{+}\rangle_{34}$ or $|\psi^{-}\rangle_{34}$)
that he has prepared initially. However, Alice's initial Bell state
($|\psi^{+}\rangle_{12}$ ) is the public knowledge so Bob knows it.
Hence, after Bell measurement is performed on qubits $2,4$ resulting
in entanglement swapping on qubits $1,3$ and $2,4$ in Eq. \ref{eq:Entanglement_Swapping},
Alice and Bob also know the initial Bell state (qubits $1,3$) shared
between them in Step 4 on which they have encoded their secret messages.
Further, they know Charlie's final Bell measurement result announcement
and their own encoding unitary operation. Therefore, they can successfully
decode each others secret information and complete the quantum dialogue
protocol between them via the help of an untrusted measurement device,
Charlie.

Now, we would like to note that there is a special reason why we have
selected that Alice prepares $|\psi^{+}\rangle_{12}$ and Bob prepares
randomly in $|\psi^{+}\rangle_{34}$ or $|\psi^{-}\rangle_{34}$ as
initial Bell state in Step 1. Specifically, in the security subsection
\ref{subsec:Security-of-OSB}, we will show how the number of initial
Bell state selection will allow Alice and Bob to reduce the information
leakage by 1-bit of classical information in comparison to the standard
quantum dialogue protocol \cite{Ba_An_QD}, where the information
leaks by 2-bits on the Bell measurement announcement. Further, we
will also show that if Alice also chooses to prepare any two Bell
states randomly in $|\psi^{+}\rangle_{12}$ or $|\psi^{-}\rangle_{12}$
like Bob chooses to prepare randomly in $|\psi^{+}\rangle_{34}$ or
$|\psi^{-}\rangle_{34}$ as initial Bell state in Step 1, then the
leakage problem can be completely solved, but in such a case, Alice
also have to securely share with Bob the exact information of the
initial Bell state $|\psi^{+}\rangle_{12}$ or $|\psi^{-}\rangle_{12}$
prepared in Step 1 as was done by Bob executing the OSB MDI-QSDC protocol
as shown in Sect. \ref{sec:OSB-MDI-QSDC}. 

Interestingly, in this way Alice and Bob can ensure that the ignorance
of Eve should be equal to the total classical information (4-bits
secret messages i.e., 2-bits by Alice and 2-bits by Bob) transmitted
between them simultaneously on the same quantum channel. Basically,
as Eve's ignorance increases, the c-bits information leakage decreases.
Therefore, the random preparation of initial Bell states by Alice
and Bob in Step 1 provides a tradeoff to avoid the conventional information
leakage problem in QD protocols. We have shown the details in the
security section. However, it's useless for Alice and or Bob to prepare
more than any two Bell states randomly for example, suppose they prepare
$|\psi^{+}\rangle_{12}$ or $|\psi^{-}\rangle_{12}$ and $|\psi^{+}\rangle_{34}$,
$|\psi^{-}\rangle_{34}$ or $|\phi^{+}\rangle_{34}$ respectively
because in this case Eve's ignorance would exceed the maximum requirements
of 4-bits in our OSB MDI-QD protocol, so we restrict ourselves to
the above two cases, however, such a case might be useful where the
Eve's ignorance requirements are higher than 4-bits.

In the original two-party QD protocols \cite{Ba_An_QD}, initial and
final Bell states are the public knowledge, and Alice and Bob know
their own encoding so they can deduce each others' encoding information.
However, there exists a partial information leakage problem of two-bits
due to the final Bell measurement announcement, i.e., Eve can extract
two-bits information about the product of unitary operations $U_{B}U_{A}$
applied by Alice $(U_{A})$ and Bob $(U_{B})$ respectively. Hence,
information leakage is the inherent problem of QD protocols. To avoid
such a leakage, we used the tricks mentioned and utilized in \cite{AQD},
such that in the first step of our OSB MDI-QD protocol, Bob can use
OSB MDI-QSDC protocol proposed in Sect. \ref{sec:OSB-MDI-QSDC} for
sharing information about his initial Bell state (which he randomly
prepared in $|\psi^{+}\rangle_{34}$ or $|\psi^{-}\rangle_{34}$).
This strategy of random preparation of Bell states by Bob would help
to increase the ignorance of Eve thereby reducing the information
leakage. 

\section{Security Analysis\label{sec:Security-Analysis}}

In this section, we have analyzed the security of the two proposed
protocols under the following possible eavesdropping attacks. 

\subsection{Security of OSB MDI-QSDC protocol}

Since Charlie is an untrusted third party (measurement device) who
is equivalent to Eve. To decode the secret message of Alice, Charlie
or Eve must have to know the initial Bell state shared between Alice
and Bob (on which Alice is supposed to encode her secret messages)
which is obtained by them after the entanglement swapping. Basically,
if the initial Bell states prepared by both Alice and Bob are known
to Charlie/Eve then after the entanglement swapping performed by Charlie,
he/she can get the knowledge of new born Bell states (qubits $1,3$)
and can decode the secret message at the end. However, Eve's attempt
to try knowing that new born initial Bell state $(1,3)$ fails because
Bob has prepared his initial Bell state randomly in one of the two
$|\psi^{+}\rangle_{34}$ or $|\psi^{-}\rangle_{34}$ Bell states (unknown
to Charlie/Eve). As soon as Charlie has performed the Bell measurement
on qubits $(2,4)$ which results into entanglement swapping on qubits
$(1,3)$ as shown in Eq. \ref{eq:Entanglement_Swapping}. After that
Charlie/Eve can never verify the exact initial Bell state shared between
Alice and Bob and they end up with the two possible product state
equally probable due to the random preparation of initial Bell state
by Bob as shown in Eq. \ref{eq:Entanglement_Swapping}. Further, any
attempts of eavesdropping by Charlie/Eve will disturb the maximal
entanglement correlation between Alice and Bob and will be traced
out by Alice and Bob, where the security would arise from monogamy
of entanglement \cite{GV_QSDC1}. All kinds of fake particle attack
or entangle-and-measure attack will be detected during the correlation
security and eavesdropping checks performed in steps 3 and 6 in Sect.
\ref{sec:OSB-MDI-QSDC}. 

\subsection{Intercept-and-resend attack}

If Eve applies an intercept-and-resend attack to our OSB MDI-QSDC,
she prepares her own Bell state, intercepts Bob\textquoteright s qubit
from the extended $B_{E}^{\prime}$ sequence during its transmission
to Charlie (in Step 2), and sends one qubit of her fake Bell state
to Charlie. Eve aims for her fake Bell state to undergo entanglement
swapping (in Step 4) with Alice\textquoteright s qubit sequence $A_{M}$,
on which Alice will later apply her secret encoding operation (in
Step 5). Since Alice's initial Bell state $|\psi^{+}\rangle_{12}$
is publicly known, Eve plans to deduce Alice\textquoteright s secret
information once Charlie announces his Bell measurement results in
Step 7. However, this attack can be detected as early as Step 3 under
Case I, when Alice and Bob perform correlation checks based on Charlie\textquoteright s
Bell measurement outcomes. Any attempted intercept-resend attack by
Eve will break the correlation, which ensures security of against
both Eve and Charlie.

\subsection{Entangle-and-measure attack}

In the entangle-and-measure attack, Eve prepares one or more ancilla
qubits, either in a pure state like $|0\rangle$ or a superposition
state $|p\rangle=\alpha|0\rangle_{e}+\beta|1\rangle_{e}$. She then
entangles these ancilla qubits with the travel qubit being transmitted
from the sender (Alice) to the receiver (Charlie) via a unitary operation,
typically a controlled-$NOT$ ($CNOT$) gate. The goal of this attack
is to allow Eve to extract information about Alice's qubits by measuring
her entangled ancilla qubits at a later stage, without directly interfering
with the main communication process. In this scenario, as Alice prepares
a decoy Bell state $|\psi^{+}\rangle_{d_{1}d_{2}}=\frac{1}{\sqrt{2}}\left(|00\rangle+|11\rangle\right)_{ht}$,
keeping the home qubit ($h$) and sending the travel qubit ($t$)
to the Charlie. Eve, with her ancilla qubit, say $|p\rangle=\alpha|0\rangle_{e}+\beta|1\rangle_{e}$.,
applies a CNOT gate, with target on the $t$ qubit and control on
her ancilla qubit $|p\rangle.$ However, after this operation, the
resulting composite state becomes separable:

\[
\begin{array}{lcl}
CNOT_{e\rightarrow t} & = & \frac{1}{\sqrt{2}}\left(|00\rangle+|11\rangle\right)_{ht}\left(\alpha|0\rangle_{e}+\beta|1\rangle_{e}\right),\\
 & = & \frac{1}{\sqrt{2}}\left[|0\rangle_{h}\alpha|00\rangle_{te}+|1\rangle_{h}\alpha|10\rangle_{te}+|0\rangle_{h}\beta|11\rangle_{te}+|1\rangle_{h}\beta|01\rangle_{te}\right],\\
 & = & \frac{1}{\sqrt{2}}\left[\left(|0\rangle+|1\rangle\right)_{h}\left(|0\rangle+|1\rangle\right)_{t}\left(\alpha|0\rangle_{e}+\beta|1\rangle_{e}\right)\right],
\end{array}
\]
which means that Eve\textquoteright s qubit is now independent from
the system. Although her eavesdropping attempt goes undetected, she
gains no useful information, as her ancilla qubit $|p\rangle$ remains
unchanged.

\subsection{Security of OSB MDI-QD protocol against information leakage\label{subsec:Leakage}}

Information leakage is an inherent aspect of QD protocols \cite{Our_QD},
quantifiable as the discrepancy between the total information transmitted
by legitimate users and the minimum amount of information Eve requires
to infer that data (i.e., Eve's ignorance). This limitation similarly
applies to the proposed OSB MDI-QD protocol, where Eve may gain some
information about Alice's and Bob's encoding once Charlie announces
the Bell measurement results. We calculate the average information
Eve may gain \cite{AQD} which can be expressed as,

\begin{equation}
I(AB:E)=H_{apriori}\text{\textminus}H_{aposteriori},\label{eq:Leakage}
\end{equation}

where, $H_{apriori}$ represents the total classical information exchanged
between Alice and Bob, which is 4 bits in our OSB MDI-QD. Now, we
calculate the Eve\textquoteright s ignorance i.e., $H_{aposteriori}$
after Charlie announces his Bell measurement outcome on qubits $1,3$
in Step 7. Let's say Charlie's Bell measurement outcome in Step 4
is $|\psi^{+}\rangle_{24}$ and in Step 7 also is $|\psi^{+}\rangle_{13}.$
This arrises two cases for Charlie/Eve considering the two random
choices of initial Bell states $|\psi^{+}\rangle_{34}$ or $|\psi^{-}\rangle_{34}$
prepared by Bob given Alice always prepares $|\psi^{+}\rangle_{12}$:

\textbf{Case (i)} When Alice and Bob have chosen to prepare $|\psi^{+}\rangle_{12}\otimes|\psi^{+}\rangle_{34}$
as initial Bell states as shown in $I$ row and $II$ column of Table
\ref{tab:The--sign}. After the entanglement swapping on qubits $2,4$
in Step 4 and encodings by Alice and Bob on qubits $1,3$ in Step
5 the state can be expressed as 
\[
|\psi^{+}\rangle_{12}\otimes|\psi^{+}\rangle_{34}=\frac{1}{2}\left(|\psi^{+}\rangle_{13}|\psi^{+}\rangle_{24}+|\phi^{+}\rangle_{13}|\phi^{+}\rangle_{24}+|\phi^{-}\rangle_{13}|\phi^{-}\rangle_{24}+|\psi^{-}\rangle_{13}|\psi^{-}\rangle_{24}\right),
\]

\begin{equation}
P(II,|\psi^{+}\rangle_{13}||\psi^{+}\rangle_{13})=P(XX,|\psi^{+}\rangle_{13}||\psi^{+}\rangle_{13})=P(iYiY,|\psi^{+}\rangle_{13}||\psi^{+}\rangle_{13})=P(ZZ,|\psi^{+}\rangle_{13}||\psi^{+}\rangle_{13})\label{eq:4_possibilities}
\end{equation}

\textbf{Case (ii)} When Alice and Bob have chosen to prepare $|\psi^{+}\rangle_{12}\otimes|\psi^{-}\rangle_{34}$
as initial Bell states as shown in $II$ row and $II$ column of Table
\ref{tab:The--sign}. After the entanglement swapping on qubits $2,4$
in Step 4 and encodings by Alice and Bob on qubits $1,3$ in Step
5 the state can be expressed as 
\[
|\psi^{+}\rangle_{12}\otimes|\psi^{-}\rangle_{34}=\frac{1}{2}\left(|\psi^{+}\rangle_{13}|\psi^{-}\rangle_{24}-|\phi^{+}\rangle_{13}|\phi^{-}\rangle_{24}-|\phi^{-}\rangle_{13}|\phi^{+}\rangle_{24}+|\psi^{-}\rangle_{13}|\psi^{+}\rangle_{24}\right),
\]

\begin{equation}
P(IZ,|\psi^{-}\rangle_{13}||\psi^{+}\rangle_{13})=P(ZI,|\psi^{-}\rangle_{13}||\psi^{+}\rangle_{13})=P(XiY,|\psi^{-}\rangle_{13}||\psi^{+}\rangle_{13})=P(iYX,|\psi^{-}\rangle_{13}||\psi^{+}\rangle_{13})\label{eq:4_possibilities-1}
\end{equation}

In both the cases above, Charlie/Eve never know that which case was
chosen by Alice and Bob. Consequently, Charlie as never been sure
about his Bell measurement outcome on qubits $2,4$ in Step 4 is coming
from Case I or Case II. So, he is also not certain which initial Bell
states of qubits $1,3$ were being shared between Alice and Bob before
encoding. Hence, all the above $8$ possibilities (encoding by Alice
and Bob) as shown together in Eq. \ref{eq:4_possibilities} and Eq.
\ref{eq:4_possibilities-1} are equally probable for Charlie/Eve with
an unknown initial Bell state of qubits $1,3$ of Case I and Case
II. Now, we can calculate $H_{aposteriori}=-8(\frac{1}{8})log_{2}(\frac{1}{8})=3$
bits. Further, $I(AB:E)=4-3=1$ bit. This is an advantage that we
have reduced Eve's ignorance by $1$ bit in comparison with the leakage
($2$ bits) in the standard QD protocol \cite{Ba_An_QD}. 

We can also consider Case (iii) $|\psi^{+}\rangle_{12}\otimes|\phi^{+}\rangle_{34}$
and Case (iv) $|\psi^{+}\rangle_{12}\otimes|\phi^{-}\rangle_{34}$
to maximize Eve's ignorance by $4$ bits and can obtain Eve's gain
$I(AB:E)=0$, which means that Eve can not obtain any information
through the public announcements of BMO by Charlie because the total
classical information exchanged between Alice and Bob is equal to
the Eve's ignorance.

\subsection{Flip Attack \label{subsec:Flip-Attack:}}

Flip attack is a kind of disturbance attack where Eve applies $X$
operation on all the travel qubits in order to misguide authorized
parties while she can not learn any meaningful information. As we
know that GV subroutine \cite{Our-OSB} fails (e.g., $X\otimes X|\psi^{+}\rangle_{dd^{\prime}}=|\psi^{+}\rangle_{dd^{\prime}}$)
when Eve tries to apply the flip attack on all the traveling qubits.
To avoid such attack in general, Alice can prepare the $n$ number
of decoy Bell pairs say in $|\psi^{+}\rangle_{dd^{\prime}}$, but
she concatenates $n-m$ Bell pairs $|\psi^{+}\rangle_{dd^{\prime}}$,
and only partner particles of the $m$ $(m<n)$ Bell pair $|\psi^{+}\rangle_{dd^{\prime}}$
in the message sequence. Specifically, $n-m$ Bell pairs will take
care of the GV subroutine to check eavesdropping and partner particles
of $m$ Bell pairs will be used for correlation check between Alice
and Bob to check flip attack. Specifically, Alice keeps the first
qubits of the $m$ Bell pairs as home qubit and sends the corresponding
second qubits to Charlie along with the message sequence in Step 5
of OSB MDI-QSDC and OSB MDI-QD in Sect. \ref{sec:OSB-MDI-QSDC} and
in Sect. \ref{sec:OSB-MDI-QD}, respectively. After Charlie announces
the receipt of all the qubits, Alice will announce the positions of
the partner particles of the $m$ Bell pairs, then Charlie measures
in $\{0,1\}$ basis and announces the results for correlation check
for Alice where Alice also measure the corresponding home qubits in
$\{0,1\}$ basis and check for the perfect correlations (if Charlie
gets 0 (1) then Alice should also get 0 (1)). If Eve really attacks
all the qubits by flip attack in the extended sequence, then the correlation
must be mismatched (i.e., if Charlie gets 0 (1) then Alice should
get 1 (0). Similarly, Bob will also check for the perfect correlation
with Charlie. Therefore, if Eve is applying flip attack on all of
the travel qubits of (both) the sequence(s) coming from Alice (and
Bob) to Charlie then she would be traced by Alice (and Bob) in correlation
check of Bell states as mentioned above. If the errors are below the
certain threshold value then Alice and Bob are safe enough to correctly
decode each-others encoding in Step 7 of OSB MDI-QD, because Eve can
not differentiate between message and decoy qubits so she will flip
all the travel qubits (messages and decoys), and applies flip operation
$X\otimes X$ on $A_{M}^{\prime},B_{M}^{\prime}$ which will never
change the actual encodings of Alice and Bob as $X\otimes X=I$, hence
no disturbance in the secret information. Similarly, Bob correctly
decodes Alice's encoding in Step 7 of OSB MDI-QSDC, as if Eve applies
flip operation $X\otimes X$ on $A_{M}^{\prime},B_{M}$ does not change
the actual encoding of Alice. Now, if Eve is applying flip attack
on few of the qubits in any of the sequences coming from Alice or
Bob to Charlie then she would be traced by them with the usual GV
subroutine. So, Eve has no way to escape without being detected.

\subsection{Disturbance attack \cite{DoS_Attack} or modification attack \cite{Modification_attack}}

This attack is a specific type of denial-of-service (DoS) attack,
where Eve aims to mislead Alice and Bob by altering the message content\textemdash such
as changing the qubit order or applying unitary transformations to
some qubits\textemdash during the transmission of the sequence $A_{M}^{\prime\prime}$
from Alice to Charlie in Step 5 of the OSB MDI-QSDC protocol, and
the transmission of sequences $A_{M}^{\prime\prime}$ and $B_{M}^{\prime\prime}$
from Alice and Bob, respectively, to Charlie in Step 5 of the OSB
MDI-QD protocol. Importantly, Eve\textquoteright s goal in this attack
is not to extract any meaningful information \cite{DoS_Attack}, but
merely to disrupt communication. However, any such manipulation will
be detectable. Since Eve cannot selectively alter only the message
qubits without affecting the decoy Bell pairs $|\psi^{+}\rangle_{dd'}$,
her interference will introduce detectable errors. These errors can
be identified in Step 6 of both protocols, ensuring the integrity
of the communication.

\section{Conclusion}

MDI-QSDC protocols are designed to eliminate security vulnerabilities
tied to imperfections in the measurement devices used in quantum communication
protocols. These vulnerabilities, such as side-channel attacks, are
common in traditional setups, where the eavesdropper can exploit weaknesses
in detectors or measurement systems. Technically, MDI-QSDC achieves
this by utilizing Bell state measurements and entanglement swapping.
Our innovative protocols utilize unique resources to effectively eliminate
security vulnerabilities associated with Charlie's measurement devices.
Additionally, they significantly enhance the range of secure direct
message transmission, achieving double the distance for secure direct
message transmission compared to conventional quantum communication
methods. We proposed two OSB protocols of MDI-QSDC and MDI-QD protocols
that are fundamentally distinct from conventional conjugate-coding
methods, offer unconditional security derived from the monogamy of
entanglement \cite{Our-OSB,GV_QSDC1}. Further, we conducted security
analysis of our protocols, evaluating their robustness against key
quantum attacks like intercept-and-resend, entangle-and-measure, flip,
and disturbance or denial-of-service attacks. We also demonstrated
the resilience of the proposed OSB MDI-QD protocol, particularly in
mitigating information leakage and ensuring the integrity of quantum
communication channels. This intrinsic security feature arises because
the correlations shared by legitimate parties exclude any potential
eavesdropper, making OSB protocols highly robust against various quantum
attacks \cite{GV_QSDC1}. Specifically, the proposed OSB MDI-QSDC
and OSB MDI-QD protocols contribute to the growing body of work in
MDI-QSDC by exploring alternative quantum mechanical resources beyond
HUP for ensuring unconditional security. These protocols demonstrate
that HUP is not the only mechanism capable of providing unconditional
security in quantum communication, opening the door for further research
into different foundational principles. Another significant motivation
is that the OSB secure quantum communication protocols outperform
under certain noisy environment in comparison to the conjugate-coding
based secure quantum communication protocols \cite{Kishore-Noise-paper}.
It remains the future prospects of our research direction that is
to investigate the scope of decoherence free subspace for the variety
of noises including non-markovian noise aligning with the recent work
\cite{DI_QSDC_Majumdar} . Moreover, it would be of wider interest
to investigate and compare the effect of various noisy channels on
our OSB MDI-QSDC and OSB MDI-QD protocol and MDI-QSDC and MDI-QD protocols
given \cite{MDI-QSDC2,MDI-QSDC-LONG}, which is under the scope of
upcoming work. 
\begin{description}
\item [{Acknowledgements}]~
\end{description}
Chitra Shukla thanks the support provided by SnT, University of Luxembourg.
This work was funded by the European Union \textendash{} Next Generation
EU, with the collaboration of the Luxembourgish Government - Department
of M\textasciiacute edia, Connectivity and Digital Policy in the framework
of the RRF program. Abhishek Shukla, Milos Nesladek acknowledge the
support from project BeQCI cofounded by EU and Belspo.

\end{document}